\documentclass[submission, Phys]{SciPost}

\usepackage[utf8]{inputenc} 
\usepackage[T1]{fontenc} 	
\usepackage[english]{babel} 

\usepackage[bitstream-charter]{mathdesign}
\usepackage{geometry} 		
\usepackage{amsmath} 		
\usepackage{mathtools} 		
\usepackage{float} 			
\usepackage{graphicx} 		
\usepackage{tabularx} 		
\usepackage{booktabs} 		
\usepackage{color, xcolor} 	
\usepackage{pdfpages} 		
\usepackage{extarrows} 		
\usepackage{multirow} 		
\usepackage{multicol} 		
\usepackage{enumitem} 		
\usepackage{xspace} 		
\usepackage{stackrel} 		
\usepackage{siunitx}
\usepackage{tikz} 			
\usepackage{braket} 		
\usepackage{bm} 			
\usepackage{tensor} 		
\usepackage{slashed} 		
\usepackage{siunitx} 		
\usepackage{lastpage} 		
\usepackage{cite} 			
\usepackage[normalem]{ulem} 
\usepackage{fontawesome} 	
\usepackage{tocloft} 		
\usepackage{titlesec} 		
\usepackage{doi} 			
\usepackage{hyperref} 		
\usepackage[most]{tcolorbox} 					
\usepackage[nameinlink, capitalize]{cleveref} 	
\usepackage[nottoc, notlot, notlof]{tocbibind} 	
\usepackage[ruled, vlined]{algorithm2e} 		
\usepackage{makecell}
\usepackage[makeroom]{cancel}
\usepackage{feynmf}
\usepackage{cancel}

\makeatletter
\def\BState{\State\hskip-\ALG@thistlm}
\makeatother

\makeatletter
\@ifundefined{pdfoutput}{}{\DeclareGraphicsRule{*}{mps}{*}{}}
\makeatother

\makeatletter
\DeclareRobustCommand*{\bfseries}{%
   \not@math@alphabet\bfseries\mathbf
   \fontseries\bfdefault\selectfont
   \boldmath
}
\makeatother

\hypersetup{
	pdftitle={Learned Uncertainties},
	pdfauthor={},
	colorlinks=true, 			
	linkcolor={red!50!black}, 	
	citecolor={blue!50!black}, 	
	urlcolor={blue!80!black} 	
} 

\DeclareSymbolFont{usualmathcal}{OMS}{cmsy}{m}{n}
\DeclareSymbolFontAlphabet{\mathcal}{usualmathcal}

\SetArgSty{textnormal}
\SetKwComment{Comment}{{\small\#}~}{}
\SetCommentSty{mycommfont}

\setitemize{itemsep=0pt, parsep=0pt} 				
\setenumerate{itemsep=0pt, parsep=0pt} 				
\setitemize{itemsep=2pt,topsep=2pt,parsep=0pt,partopsep=0pt,leftmargin=*}
\setenumerate{itemsep=0pt,topsep=2pt,parsep=0pt,partopsep=0pt,labelindent=3pt,leftmargin=*}
\newlist{todolist}{itemize}{2}
\setlist[todolist]{label=$\square$}
\usepackage{pifont}

\usepackage{amsmath}
 
\usepackage{amsthm} 		
\theoremstyle{definition}

\usetikzlibrary{arrows, shapes.geometric, arrows.meta, shapes, decorations.pathreplacing, fit, patterns, patterns.meta}

\definecolor{Rcolor}{HTML}{E99595}
\definecolor{Gcolor}{HTML}{C5E0B4}
\definecolor{Bcolor}{HTML}{9DC3E6}
\definecolor{Ycolor}{HTML}{FFE699}
\definecolor{Ycolor_light}{HTML}{FFF7DE}
\definecolor{Gcolor_light}{HTML}{F1F8ED}

\tikzstyle{expr} = [circle, minimum width=1.8cm, minimum height=1.8cm, text centered, align=center, inner sep=0, draw,font=\LARGE]
\tikzstyle{txt_huge} = [align=center, font=\Huge, scale=2]
\tikzstyle{txt} = [align=center, font=\LARGE, minimum height=1cm]
\tikzstyle{cinn} = [double arrow, double arrow head extend=0cm, double arrow tip angle=130, shape border rotate=90, inner sep=0, align=center, minimum width=2.1cm, minimum height=2.3cm, fill=Bcolor, draw,font=\LARGE]
\tikzstyle{cinn_black} = [cinn, minimum height=2.5cm, fill=black]
\tikzstyle{arrow} = [thick,-{Latex[scale=1.0]}, line width=0.2mm, color=black]
\tikzstyle{loss} = [rectangle, align=center,  minimum width=1.8cm, minimum height=1.5cm,fill=Rcolor,font=\LARGE, rounded corners]
\tikzstyle{xt} = [rectangle, align=center,  minimum width=5cm, minimum height=1.5cm,fill=Gcolor,font=\LARGE, rounded corners]
\tikzstyle{xts} = [rectangle, align=center,  minimum width=1cm, minimum height=1.5cm,fill=Gcolor,font=\Large, rounded corners]

\tikzstyle{embed} = [rectangle, rounded corners=0.3ex, minimum width=1.5cm, minimum height=1cm, text centered, align=center, inner sep=0, fill=Ycolor, font=\large, draw]

\tikzstyle{small_cinn} = [double arrow, double arrow head extend=0cm, double arrow tip angle=130, inner sep=0, align=center, minimum width=1.1cm, minimum height=0.5cm, fill=Bcolor, draw]

\tikzstyle{transformer} = [rectangle, rounded corners, minimum width=6cm, minimum height=2.4cm, font=\large, fill=Gcolor_light, draw]

\tikzstyle{attention} = [rectangle, rounded corners=0.3ex, minimum width=5.5cm, minimum height=1.2cm, align=center, fill=Gcolor, draw, font=\large] 

\definecolor{red_cb}{HTML}{e41a1c}
\definecolor{blue_cb}{HTML}{377eb8}
\definecolor{green_cb}{HTML}{4daf4a}
\definecolor{purple_cb}{HTML}{984ea3}
\definecolor{orange_cb}{HTML}{ff7f00}

\definecolor{EmeraldGreen}{HTML}{1ea78d}
\definecolor{EnglishRed}{HTML}{b02427}
\hypersetup{colorlinks=true,urlcolor=EmeraldGreen,citecolor=EmeraldGreen,linkcolor=EnglishRed}

\newcommand{\angstrom}{\text{\normalfont\AA}}

\newcommand\one{\leavevmode\hbox{\small1\normalsize\kern-.33em1}}

\newcommand{\arXiv}[2][]{%
	\ifthenelse{\equal{#1}{}}%
	{\href{http://arxiv.org/abs/#2}{arXiv:#2}}%
	{\href{http://arxiv.org/abs/#2}{arXiv:#2~[#1]}}}

\def\slashchar#1{\setbox0=\hbox{$#1$}           
   \dimen0=\wd0                                 
   \setbox1=\hbox{/} \dimen1=\wd1               
   \ifdim\dimen0>\dimen1                        
      \rlap{\hbox to \dimen0{\hfil/\hfil}}      
      #1                                        
   \else                                        
      \rlap{\hbox to \dimen1{\hfil$#1$\hfil}}   
      /                                         
   \fi}

\newcommand{\tikznode}[2]{%
\ifmmode%
\tikz[remember picture,baseline=(#1.base),inner sep=0pt] \node (#1) {$#2$};%
\else
\tikz[remember picture,baseline=(#1.base),inner sep=0pt] \node (#1) {#2};%
\fi}

\def\mathswitchr#1{\relax\ifmmode{\text{#1}}\else$\text{#1}$\xspace\fi}
\def\mathswitch#1{\relax\ifmmode#1\else$#1$\xspace\fi}

\graphicspath{{./Figures/}}

\begin{document}

\begin{center}{\Large \textbf{
Towards Precise Simulations and Inference for the Neutron EDM
}}\end{center}

\begin{center}
  Skyler Degenkolb\textsuperscript{1}, 
  Luigi Favaro\textsuperscript{2}, 
  Peter Fierlinger\textsuperscript{5},  
  Jennifer Franz\textsuperscript{5}, \\
  Husain Manasawala\textsuperscript{1,5}, and
  Tilman Plehn\textsuperscript{3,4}
\end{center}

\begin{center}
{\bf 1} Physikalisches Institut, Universit\"at Heidelberg, Germany\\
{\bf 2} CP3, Universit\'e catholique de Louvain, Louvain-la-Neuve, Belgium\\
{\bf 3} Institut f\"ur Theoretische Physik, Universit\"at Heidelberg, Germany\\
{\bf 4} Interdisciplinary Center for Scientific Computing (IWR), Universit\"at Heidelberg, Germany \\
{\bf 5} TUM School of Natural Sciences,
Technische Universität München 
\end{center}



\section*{Abstract}
          Precision measurements of neutron properties, like its permanent electric dipole moment, rely on understanding complex experimental setups in detail. We show how the properties of stored and transported ultracold neutron ensembles can be simulated reliably. In a second step, we illustrate how  they can be used for simulation-based inference of the parameters associated with underlying physics processes such as neutron capture or beta decay. Our proof of principle for simulation-based inference confronts a longstanding challenge with ultracold neutrons: low measurement statistics coupled with a complex apparatus. 

\vspace{1pt}
\noindent\rule{\textwidth}{1pt}
\tableofcontents\thispagestyle{fancy}
\noindent\rule{\textwidth}{1pt}

\clearpage
\section{Introduction}
\label{sec:intro}

Ultracold neutrons (UCN) have kinetic energies sufficiently low to be trapped and manipulated in experiments for durations of many seconds or minutes, enabling a variety of precision measurements in low-energy particle physics.
The neutron lifetime and permanent electric dipole moment (EDM) are, respectively, key pieces of evidence for understanding the abundance of elements in the early universe and the observed over-abundance of matter relative to antimatter~\cite{Dubbers:2011ns,ParticleDataGroup:2024cfk}.
The neutron EDM has a long and distinguished history of increasingly precise null results~\cite{Chupp:2017rkp}, of which those obtained from UCN are by far the most precise~\cite{Abel:2020pzs}.

Via the EDM, UCN deliver one of the most precise tests of the violated fundamental symmetries $P$ and $T$, and stringently constrain $CP$-violating physics beyond the Standard Model~\cite{Degenkolb:2024eve}.
The finite neutron lifetime $\tau_n \approx 880\,$s is also a key ingredient to determine Standard Model couplings, such as the CKM matrix element $V_{ud}$ or the ratio $\lambda=g_A/g_V$ of the hadronic weak axial and vector couplings.
The value of $\tau_n$ obtained with UCN~\cite{Musedinovic:2024gms,Serebrov:2017bzo}, while significantly more precise, is also significantly discrepant with that measured using neutron beams~\cite{Yue:2013qrc,Byrne:1996zz}.

Other measurements exploiting UCN in particle physics include angular correlations in $\beta$ decay~\cite{UCNA:2012fhw}, bound states in Earth's gravity~\cite{Cronenberg:2018qxf,Jenke:2020obe}, tests of Lorentz invariance~\cite{Altarev:2009wd,Ivanov:2020son}, searches for axion-like new particles ~\cite{Ayres:2023txi}, and limits on the oscillation of neutrons to other neutral particles~\cite{Ban:2023cja}.
In general, UCN have provided an advantage in experiments where the benefits of long measurements outweigh the disadvantages of low statistics.
UCN experiments are typically statistics-limited, and the development of improved sources has become a major preoccupation of the field.
While we focus here on the specific science case of EDM experiments served by a helium-based superthermal UCN source, our approach and methods are general.

Except for the special case of fully \textit{in-situ} measurements~\cite{doi:10.3233/JNR-220044}, UCN must be extracted from a source for delivery to experiments.
Delivery efficiency is typically on the percent-level or less, and in addition can depend strongly on UCN energy.
While UCN storage and transport can be modeled analytically with some success, simple models do not permit a precise confrontation with experimental data.
Monte Carlo simulations are challenging and computationally intensive, and it is not straightforward to determine which of many correlated parameters ultimately drives experimentally observed variations.
In addition, data for UCN measurements come only from counting neutrons at the end of long and complex experimental sequences, without the possibility for online monitoring or guided intervention.
It is therefore of great interest to develop methods that shed light on the intermediate processes, ideally with quantified uncertainties for physically meaningful parameters.

Driven by modern machine learning developments~\cite{Plehn:2022ftl} the availability of precise first-principle simulations is closely tied to simulation-based inference (SBI)~\cite{Cranmer:2019eaq}, for optimal analyses of complex experimental setups~\cite{Dax:2021tsq,Dax:2023ozk}. Here, SBI techniques are making rapid technical progress and by now go much further than extracting fundamental parameters from legacy datasets. Generative methods~\cite{Bellagente:2019uyp,Bellagente:2020piv,Bieringer:2020tnw} also allow us to control correlated high-dimensional vectors of technically motivated experimental parameters or to unfold data to an representation that allows, for instance, for an efficient combination of different analyses.

In Sec.~\ref{sec:ucn} we given an overview of pertinent UCN physics, before in Sec.~\ref{sec:vTOF} we present the first precision simulations of \textit{in-situ} UCN storage and extraction from a high-density source. This forward simulation allows for the first SBI application in UCN physics, which we present in Sec.~\ref{sec:inf}.

\section{UCN production, storage, and EDM measurements}
\label{sec:ucn}

\begin{figure}[b!]
    \centering
    \includegraphics[width=0.475\textwidth]{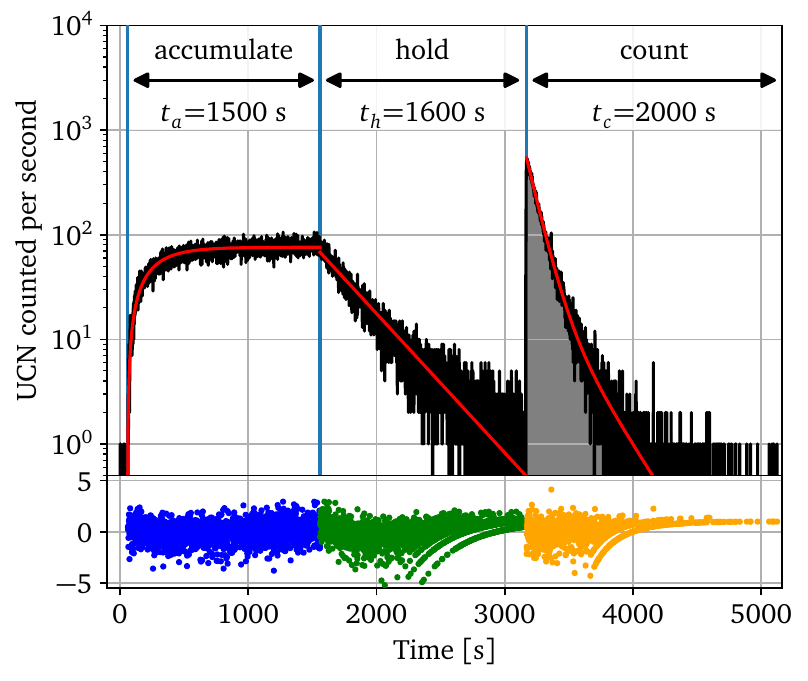}
    \includegraphics[width=0.515\textwidth]{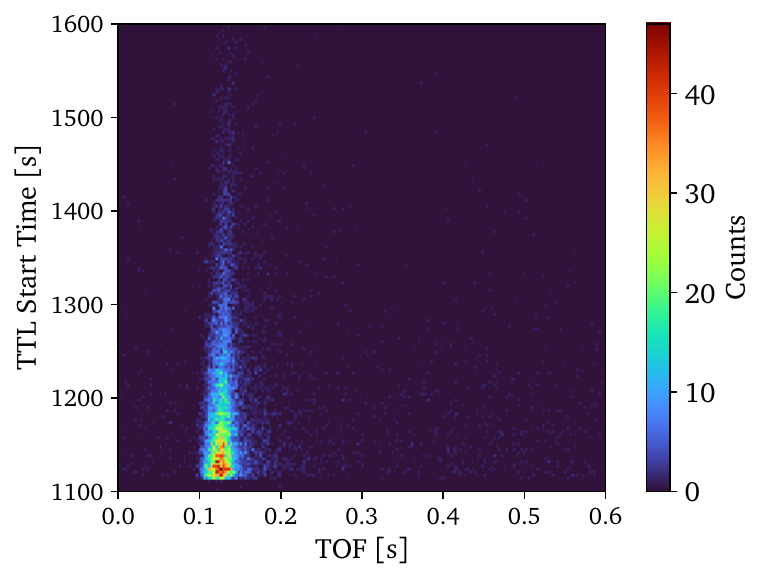}
    \caption{\textbf{Left:} Integral UCN measurement, showing the production and extraction sequence with phenomenological fits following~\cite{Degenkolb:2025fbc}. \textbf{Right:} Example of an experimental vTOF measurement, where $(t_a,t_h) = (1000\,\text{s},100\,\text{s})$ and UCN extraction/counting begins at $1100\,$s. The measured time-of-flight (TOF) spectra are shown as horizontal rows of colored pixels, which evolve towards later counting times along the vertical axis as UCN drain from the source. TTL start time gives the global timestamp of a trigger signal defining each chopper \textit{frame}, i.e., the zero for local timestamps within the $\sim0.6\,$s repetition period of successive TOF spectra.
    }
    \label{fig:SuperSUN_accumulation}
\end{figure}

We focus on superthermal production of ultracold neutrons (UCN), using isotopically pure superfluid-$^4$He as a conversion medium.
Unlike $^4$He, other production media rapidly absorb UCN and are used exclusively to feed external experiments.
UCN production with superfluid-$^4$He provides a unique opportunity to perform \textit{in-situ} experiments, as well as external ones.
UCN storage in this environment also enables studies of fundamental UCN interactions within the source, applying the same methods needed for external experiments.
Data from a simple \textit{in-situ} storage experiment are shown in Fig.~\ref{fig:SuperSUN_accumulation}, where UCN are released to an external detector after \textit{in-situ} accumulation and holding.
A small leakage to the external detector during accumulation and holding permits observing, respectively, the build-up and decay of the stored population~\cite{Degenkolb:2025fbc}.
Fig.~\ref{fig:SuperSUN_accumulation} also shows data measured externally by \textit{vertical time-of-flight} (vTOF), following a similar preparation sequence.
This method provides access to partial information about the stored UCN spectrum, and is the focus of our simulations and inference for this study.

Cold neutrons with $8.9\,$Å are delivered in a beam, enter a cryostat, and downscatter in a superfluid-filled trap. The dominant production channel is inelastic scattering where a single phonon is produced, and the neutron imparts nearly all its energy to the helium. The resulting UCN have energies $10^5\times$ lower than the original cold neutrons, and can be stored in closed containers (such as the converter itself) for holding times on the order of the neutron's $\beta$-decay lifetime, $\tau_n \approx 880\,$s.
Use of a material trap introduces additional loss from neutron capture on the confining walls, which can be minimized but not eliminated through careful choice of materials.
Presence of $^3$He impurities in the bulk of the conversion medium also causes loss, which is minimized by isotopic purification of the superfluid.
Both of these loss mechanisms tend reduce storage lifetimes, but although they impact stored UCN spectra differently, their different physical origin is difficult to resolve from experimental data such as those shown in Fig.~\ref{fig:SuperSUN_accumulation}.

\subsubsection*{UCN Production}
\label{sec:UCN_prod}

The state-of-the-art for UCN production with $^4$He is the SuperSUN instrument at the Institut Laue-Langevin~\cite{Degenkolb:2025fbc}, which we take as the concrete experimental setup for this work.
SuperSUN itself can operate in two modes: accumulation mode (as shown in Fig.~\ref{fig:SuperSUN_accumulation}), where a high UCN density is accumulated over times on the order of $100-1000\,$s to be later released in a burst, and continuous (or open converter) mode, where UCN are simultaneously produced and extracted.
The energy spectrum of UCN \textit{production} is essentially steady in time, but the stored UCN energy spectrum evolves in time due to energy-dependent loss processes.
In continuous mode a steady-state is reached relatively rapidly, due to the large converter loss associated with UCN extraction.
In accumulation mode while the converter is closed, converter losses can be very small and the energy spectrum is significantly influenced by the experimentally chosen time intervals for accumulation and holding, respectively $t_a$ and $t_h$.

UCN are detected by extracting them from the source, allowing the UCN gas to stochastically follow a guide system that leads to an external detector.
The extraction and detection processes also involve energy-dependent loss, introducing additional parameters and further shaping the detected spectrum.
Integral measurements such as that shown at left in Fig.~\ref{fig:SuperSUN_accumulation} can be performed by simply counting UCN, without intervening apparatus to distinguish different spectral components.
Time-of-flight (TOF) measurements, such as that shown on the right in Fig.~\ref{fig:SuperSUN_accumulation}, resolve one component of velocity and offer access to partial information from the UCN energy spectrum.
However, the presence of correlations; mixing of velocity components during TOF; and strong shaping of the (three-dimensional) velocity spectrum by upstream components present further challenges for reconstructing the situation within the source, or even slightly upstream of the detector.

The differential spectrum for UCN production (per unit volume, time, and energy), via the inelastic scattering process $E_0 \rightarrow E$ is
\begin{align}
    \frac{dN}{dE d^3\bm{x} dt} &= \int dE_0 \frac{d\Sigma_s\left( E|E_0 \right)}{dE}\frac{d\Phi(E_0)}{dE_0} \; ,
\end{align}
where $E_0\gg E$ and $\Sigma_s$ is the corresponding (macroscopic) cross-section with units of $\text{cm}^{-1}$ and $\Phi$ is the cold-neutron flux with units of $\text{cm}^{-2} \, \text{s}^{-1}$. In SuperSUN the most important production process is a dominating contribution from the single-phonon channel for cold neutrons with $E_0 \approx 1\,\text{meV}$. The UCN production rate is given, at leading order, by uniform filling of the phase space $d^3\bm{x}d^3\bm{k}$. The differential energy spectrum then scales as $ |\bm{\mathrm{k}}|^2d|\bm{\mathrm{k}}| \propto \sqrt{E}dE$~\cite{GOLUB1977337}, up to a cutoff energy set by the potential-energy barrier of the converter wall. Using measured values for $\Sigma_s$ and $E_0 = \frac{1}{2}m_n v_0^2 = h^2/2 m_n \lambda_0^2$ where $\lambda_0 = 8.9\,\text{\angstrom}$ is related to $v_0 \approx 440\,\text{m}/\text{s}$ by the usual de Broglie relation, the produced UCN spectrum integrated up to a maximum energy $E_\text{max}$ can be expressed as \cite{SCHMIDTWELLENBURG2009259}
\begin{align}
    \frac{dN}{d^3\bm{x} dt} &\approx 4.97(38)\times 10^{-8} \frac{\angstrom}{\text{cm}} \left. \frac{d\Phi}{d\lambda} \right|_{\lambda_0 = 8.9\,\angstrom} \left( \frac{E_\text{max}}{233\,\text{neV}} \right)^{\frac{3}{2}} \notag \\
    &\equiv C \cdot  E_\text{max} ^{\frac{3}{2}},
\label{eq:intspec}    
\end{align}
where for SuperSUN $d\Phi/d\lambda \approx 2.7\times10^8 \, \text{cm}^{-2} \, \text{s}^{-1} \,\angstrom^{-1}$ can be absorbed in the constant $C$, in the approximation that the cold neutron beam is spatially uniform within the converter.
In any case, stored UCN rapidly reach a mechanical equilibrium in which the phase space available to them becomes uniformly filled.
(Note that neutron-neutron interactions are too rare to be measured with free neutrons, and therefore thermalization within the stored ensemble effectively does not occur.)

Sub-leading production channels are also well known~\cite{SCHMIDTWELLENBURG2009259}, and represent approximately a $10\%$ contribution to total UCN production at SuperSUN. For completeness, the differential energy spectrum for production is correspondingly
\begin{align}
    \frac{dN}{dE d^3\bm{x} dt} \approx C \cdot  \frac{3}{2} \sqrt{E},
    \label{eq:diffspec}
\end{align}
and we now interpret $C$ as a generic normalization factor.

The produced UCN are trapped, up to a limiting energy $V$, by interactions with the nuclei of material walls.
This interaction is described by an effective potential barrier, the so-called neutron optical potential for that material.
For example, large sections of the converter's confining walls have $V=115\,\text{neV}$.
Superfluid helium also has a neutron-optical potential of $18.5\,\text{neV}$, implying that for wall interactions under helium, the effective trapping barrier is reduced.
In practice, the converter wall can confine neutrons up to a kinetic energy of $E_\text{max} \approx V-18.5\,\text{neV} = 96.5\,\text{neV}$.
Neutrons exiting the helium into vacuum are boosted in the direction normal to the interface, gaining $18.5\,\text{neV}$ of kinetic energy.

\begin{figure}[t]
    \includegraphics[width=0.485\textwidth]{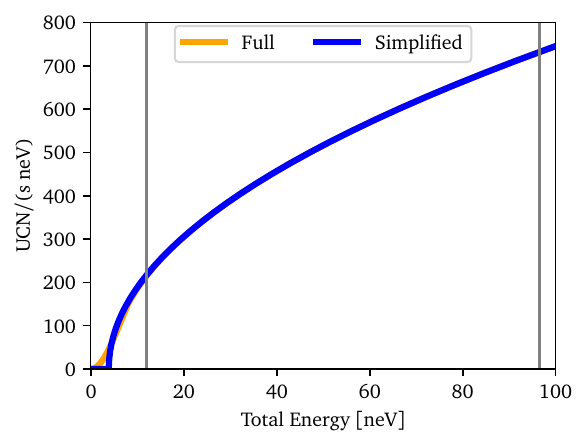}
    \includegraphics[width=0.505\textwidth]{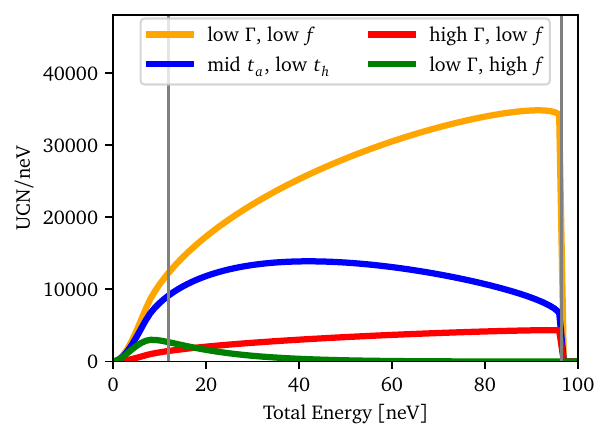}
    \caption{Total-energy UCN spectra \textit{in-situ}, calculated for SuperSUN's converter properties. The trappable \textit{and} extractable energy range lies between the gray grid-lines. \textbf{Left:} energy-dependent UCN production rate (Full: closed-form integral Eq.\eqref{eq:full_prod_spec} including all geometrical and gravitational effects / Simplified: kinetic energy spectrum shifted by the mean gravitational potential in the converter, Eq.\eqref{eq:extractable_prod_spec}). \textbf{Right:} Stored UCN spectra obtained from Eq.\eqref{eq:full_spec} for a variety of preparation sequences and converter loss rates, illustrating the impact of these parameters.
    }
    \label{fig:supersun_production_spectra}
\end{figure}

%
UCN produced with kinetic energy $E$ thus have, in addition, potential energy both from Earth's gravity and from the superfluid helium production medium. This can be extended further to magnetic fields; here we limit ourselves to the zero-field case.
A neutron in Earth's gravity has $m_ng=102.5\,\text{neV}/\text{m}$.
For convenience we set the zero of potential to be at the vertically lowest point inside the converter, within the superfluid helium.
UCN with kinetic energy $E=0$ at this point thus also have vanishing total energy.
The total energy $E_\text{tot}$ is assumed to be conserved, since inelastic processes add so much energy that the neutron becomes untrapped; such processes are therefore counted as loss channels.
The ensemble properties of produced and stored UCN at mechanical equilibrium can be calculated from the total energy~\cite{Pendlebury:1994he}.

In SuperSUN UCN are produced, and can be trapped and stored, within a cylindrical converter volume approximately $3\,$m long with $D=74.4\,$mm diameter (we neglect deviations from a cylinder in the present work, e.g., near the end where UCN are extracted~\cite{Degenkolb:2025fbc}).
Gravity acts transversely to the cylinder axis, and the total energy spectrum that results from integrating over the converter volume can be calculated analytically.
The result is equivalent to multiplying the spectrum of Eq.\eqref{eq:diffspec} by the real part of the ordinary hypergeometric function $\text{Re} [_2F_1(-\frac{1}{2},\frac{3}{2},3,\frac{m_ngD}{E})]$.

For detection or delivery to experiments, UCN are extracted vertically into a guide with diameter $d=50\,\text{mm}$.
A vertical distance $h=28.5\,$cm separates the converter's central axis from that of a horizontal guide (also $d=50\,\text{mm}$) that ultimately brings UCN out of the cryostat.
Thus, only UCN with total energy
\begin{align}
    E_\text{tot} > E_\text{min} 
    = \left(h+\frac{D}{2}-\frac{d}{2}\right)m_ng - 18.5\,\text{neV} 
    \approx 12\,\text{neV}
\end{align}
can be detected via extraction. As shown in Fig.~\ref{fig:supersun_production_spectra}, the differential total-energy spectrum above this value (i.e., what can actually be measured or used in external experiments) is indistinguishable from a shifted \textit{kinetic}-energy spectrum, Eq.\eqref{eq:diffspec} with $E\rightarrow E-\frac{m_ngD}{2}$. For measurements based on extracted UCN, we can thus approximate the \textit{in-situ} spectrum by making this replacement in Eqs.\eqref{eq:intspec} and \eqref{eq:diffspec}. The rate of total \textit{measurable} UCN production is thus
\begin{align}
    \frac{dN}{dt} &= \frac{3C}{2}\cdot \mathcal{V} \int_{E_\text{min}}^{E_\text{max}} dE_\text{tot} \sqrt{E_\text{tot}-\frac{m_ng D}{2}} 
    \notag \\
    &= C\cdot \mathcal{V}\left[ \left( E_\text{max}-\frac{m_ng D}{2}  \right)^\frac{3}{2} - \left( E_\text{min}-\frac{m_ng D}{2}  \right)^\frac{3}{2} \right] \; ,
\end{align}
where $\mathcal{V}$ is the converter volume.

It is, however, necessary to evaluate storage, loss, and extraction with energy-dependence preserved.
For this purpose, we preserve the complete total energy spectrum down to $E_\text{tot}=0$.
This enables studying the properties or UCN that cannot be extracted, but may be of interest for \textit{in-situ} experiments, without revising our methods.
It is also noteworthy that, as Fig. \ref{fig:supersun_production_spectra} illustrates, for certain preparation sequences and parameter values a significant fraction of the stored spectrum has $E_\text{tot}<E_\text{min}$.

\subsubsection*{UCN Losses}

UCN are lost via many channels, which can be broadly categorized according to whether or not they depend on UCN energy.
In general the loss probability can depend not only on energy, but also, e.g., on the angle of incidence for wall interactions.
This angular dependence is typically neglected, as is the slight anisotropy for UCN production, by assuming that UCN velocities are rapidly randomized by nonspecular reflections inside the converter.
Simulations confirm that in SuperSUN, this approach to mechanical equilibrium occurs in much less than one second. 

The total loss rate is composed of individual loss rates $\Gamma_i$ for different channels,
\begin{align}
    \frac{1}{\tau(E_\text{tot})} = \sum_i \Gamma_i = \Gamma(E_\text{tot}) + \Gamma'   \; ,
\end{align}
where $\tau(E_\text{tot})$ is the mean survival time for UCN with energy $E_\text{tot}$, and $\Gamma'$ represents the sum of all energy-independent losses.
Energy-dependent losses $\Gamma(E_\text{tot})$ arise mainly from UCN dynamics, either via the energy dependence of neutron capture or scattering at walls, or from the (velocity-dependent) frequency of such interactions.
Additional contributions to $\Gamma(E_\text{tot})$ include mechanical gaps, through which neutrons escape confinement, and higher-order scattering from quasi-particles in the superfluid helium.

Energy-dependent losses include $\beta$ decay with $\Gamma_\beta = \tau_n^{-1}\approx 0.00114\,\text{s}^{-1}$, and for fixed-duration experiments, neutron capture in the bulk of the storage medium. For fixed-\textit{length} experiments, such as TOF, a velocity-dependence re-emerges.
For our case, $^3$He capture due to trace impurities in the $^4$He production medium gives $\Gamma_{^3\text{He}} \approx x\cdot 2.4\times 10^7 \,\text{s}^{-1}$~\cite{nEDM:2019qgk}, where $x$ is the fraction of $^3$He relative to $^4$He.
Like neutron capture in the bulk, some upscattering losses also exhibit energy-independence (we recall that essentially any inelastic interaction will lead to UCN gaining energy and being lost).
The dominant mechanism of this type is two-phonon scattering, with $\Gamma_2 \approx 0.01 ( T/\text{K} )^7 \,\text{s}^{-1}$, is negligible in comparison to $\beta$ decay at SuperSUN's operating temperature or $0.6\,$K.

The dominant, and most readily calculable, contribution to $\Gamma(E_\text{tot})$ is from neutron capture or upscattering on walls.
Noting that $E = E_\text{tot} - m_n gz$, where we assume that UCN are immersed in helium within the vertical range $0<z<D$, the general expression for the wall-loss rate at mechanical equilibrium is~\cite{Pendlebury:1994he}
\begin{align}
    \Gamma_\text{wall}(E_\text{tot}) = 
    \frac{1}{4} \sqrt{\frac{2E_\text{tot}}{m_n}}
    \frac{\int_0^{z_\text{max}}\left(1-\frac{m_ngz}{E_\text{tot}}\right)\bar\mu(E_\text{tot}-m_ngz)dA(z)} {\int_0^{z_\text{max}}\sqrt{1-\frac{m_ngz}{E_\text{tot}}}d\mathcal{V}(z)}\; ,
    \label{eq:full_wall_loss}
\end{align}
where $z_\text{max} = \min(D,\frac{E_\text{tot}}{m_ng})$.
The integration measures $d\mathcal{V}(z)$ and $dA(z)$ are the differential volume and surface area, respectively, of the trap and its wall at height $z$.
The numerator represents the total wall-loss rate for an ensemble of UCN each having energy $E_\text{tot}$, accounting for both the variation of UCN density and kinetic energy with height, and the kinetic energy dependence of the mean loss probability per wall interaction $\bar\mu(E)$ (which depends also on properties of the wall, see below).
The integral in the denominator normalizes this to the total number of UCN with energy $E_\text{tot}$, such that the rate $\Gamma_\text{wall}(E)$ can be used at the particle level.

As an auxiliary quantity, the mean free path in the trap (here, the UCN converter vessel) is useful:
\begin{align}
    \Lambda \approx \frac{4\mathcal{V}}{A}\approx 73.5\,\text{mm} \; ,
\end{align}
where $A$ is the total surface area.
This expression holds, even in the presence of gravity, for UCN with $E_\text{tot}>m_n g D$ when the trap has a horizontal plane of reflection symmetry (e.g., for a horizontally oriented cylinder).
The case for $E_\text{tot}<m_n g D$ is also calculable~\cite{Pendlebury:1994he} in similar fashion to Eq.\eqref{eq:full_wall_loss}, and in our case tends to reduce $\Lambda$.

For UCN with $E_\text{tot}\approx E\gg m_n g D$, the mean wall interaction frequency is $v/\Lambda$ where we take $v$ as the mean speed within the trap (still for fixed energy).
The more commonly used expression for the wall loss rate,
\begin{align}
 \Gamma_\text{wall}(E) \sim \frac{\bar\mu(E)}{\Lambda}\sqrt{\frac{2E}{m_n}} \; ,
\end{align}
amounts to neglecting the difference between total energy and kinetic energy, and taking $\bar\mu$ outside the integral.

Useful calculations require an explicit form for $\bar\mu(E)$, and reflection from walls is treated as a simple 1D quantum mechanics problem.
The complex neutron optical potential $U=V\cdot(1-if)$ encodes loss via the dimensionless real parameter $f$ as a (typically small) correction to the real potential $V$.
For a homogeneous material in which the neutron wave vector is $\bm{k}$, which may be complex, $f=\sigma_l \text{Re} [k]/(4\pi b)$ is related to the microscopic loss cross section $\sigma_l$ and bound coherent scattering length $b$.

The physically relevant quantity for calculating reflections is the \textit{difference} of optical potential at the interface going from material 1 into material 2, i.e., $\Delta U = U_2-U_1$.
The converter wall coatings with a $115\,\text{neV}$ optical potential have $f\lesssim3\times10^{-5}$~\cite{Neulinger:2022oue}.
The loss parameter $f$ is largely unaffected for UCN being trapped or transported in non-absorbing media, but the shift of $\text{Re} \left[ \Delta U \right]$ can be significant (e.g., $16\%$ for this material in helium), resulting also in substantially modified reflection losses.

To a first approximation, physical gaps can be considered as a fully absorbing fraction of the total surface area $a$, with the ensemble- and trap-averaged consequence of adding a further $f'\approx a/A$ to the loss factor $f$ of the material itself.
This treatment neglects local geometry-dependent "trapping" effects that can be captured in simulations, i.e., local corrections to the mean free path for UCN that may return to the trap after entering a gap.

The UCN loss probability on reflection from a wall in vacuum depends on the incidence angle at the surface, which we write as $\theta$ with normal incidence for $\theta=0$. In terms of the auxiliary dimensionless variable $u_\perp(\theta)=(E/V) \cos^2\theta$, the survival probability \textit{amplitude} is
\begin{align}
   R(\theta) &= \frac{\sqrt{u_\perp(\theta)}-\sqrt{u_\perp(\theta)-1+if}}{\sqrt{u_\perp(\theta)}+\sqrt{u_\perp(\theta)-1+if}}
\end{align}
so that the survival \textit{probability} per wall interaction is
\begin{align}
   1 - \mu(E,\theta,f) &= \frac{u_\perp(\theta) + \alpha(\theta) - \sqrt{u_\perp(\theta)}\sqrt{2\alpha(\theta)-2\left( 1-u_\perp(\theta) \right)}}{u_\perp(\theta) + \alpha(\theta) + \sqrt{u_\perp(\theta)}\sqrt{2\alpha(\theta)-2\left( 1-u_\perp(\theta) \right)}} \notag \\ 
   &=\left| R(\theta) \right|^2,
   \label{eq:full_loss}
\end{align}
where $\alpha(\theta) = \sqrt{f^2 + \left( 1-u_\perp(\theta) \right)^2}$. The \textit{mean} wall loss probability per interaction is then obtained from kinetic theory, 
\begin{align}
    \bar\mu(u,f) &= \frac{\int \mu(u,\theta,f)\cos\theta d\Omega}{\int \cos\theta d\Omega}  \approx 2f \left( \frac{\sin^{-1}\sqrt{u}}{u} - \sqrt{\frac{1}{u}-1} \right) ,
   \label{eq:mu_approx}
\end{align}
where $u= E/ V$ and for clarity we now explicitly give the loss parameter $f$ as an additional argument, writing $\bar\mu(u,f)$ rather than $\bar\mu(E)$.
The right-hand side is only the leading approximation for $f\ll 1$ to a lengthy closed-form~\cite{Steyerl:2020xpc}.
While simulations typically work directly with Eq.\eqref{eq:full_loss}, calculations can employ either a full angle-averaged analytic solution or numerically-integrated look-up functions (which are rather computationally faster).
Because UCN can undergo many thousands of wall interactions in a storage experiment, small imprecisions can build up and it is desirable to avoid any but the most robust approximations.

The replacement $V\rightarrow \text{Re} \left[ \Delta U \right]$ is to be understood for reflections in which one medium is not vacuum, and for analytic calculations of real traps that incorporate multiple materials, it may be necessary to consider an effective $f$ that represents an average within the phase space accessible to the trapped ensemble.
For experiments involving a storage phase (i.e., in our treatment below), what matters is the minimum value of $V$ on the accessible wall surface.
It is to be understood in the following equations as a scaling parameter that defines the boundary between trapped and untrapped UCN, arising from the definition of $u$.

\subsubsection*{UCN Storage}

When UCN are being stored, without further production, each distinct $E_\text{tot}$ has a time-independent total loss rate given by the sum of all energy-dependent and energy-independent partial loss rates.
Now expressed in terms of $u$ rather than $E_\text{tot}$, the simplified expression for loss during this holding phase is
\begin{align}
    \Gamma_\text{hold}(u) =\tau_\text{hold}^{-1}(u) 
    \approx \frac{\bar\mu(u,f_\text{hold})\sqrt{u}}{\Lambda}\sqrt{\frac{2V}{m_n}} + \Gamma'_\text{hold} \; ,
    \label{eq:loss_hold}
\end{align}
where $\tau_\text{hold}$ is the time-constant that characterizes exponential decay for this energy class.
The probability of UCN survival then follows an exponential decay law from a fixed initial population toward zero, $\exp ( -t/\tau_\text{hold}(u) )$.
A similar equation for loss can be written for the accumulation phase, during which the loss rates might conceivably differ from those during storage, e.g., due to beam-induced heating.
Allowing for possibly different loss processes as compared to a storage scenario, we write the time constant for accumulation as $\tau_\text{acc}(u)$ -- but conceptually this works in the same way as for Eq.\eqref{eq:loss_hold}.
Each UCN energy $u$ thus has a different time constant for its approach to saturated equilibrium, which follows a function $1-\exp\left( \frac{-t}{\tau_\text{acc}(u)} \right)$. These functions can be derived by solving a rate equation, written for fixed $u$.

UCN production takes place only while the cold-neutron beam is supplied, while UCN loss takes place continuously at all times. The approach to a finite steady-state population, where the production and loss rates are equal, is in some sense the inverse process to UCN loss during storage (when new UCN are not produced, and the steady state occurs for vanishing population).
The saturated equilibrium for each energy $u$ is given by the ratio of production rate to loss rate, i.e.,
\begin{align}
    \frac{dN_\textbf{sat}}{du} = \frac{3C V^\frac{3}{2}}{2}\mathcal{V} \cdot \tau_\text{acc}(u) \sqrt{u-u_0} \; ,
    \label{eq:extractable_prod_spec}
\end{align}
or
\begin{align}
    \frac{dN_\textbf{sat}}{du} = \frac{3C V^\frac{3}{2}}{2}\mathcal{V} \cdot \tau_\text{acc}(u) \sqrt{u} \cdot \text{Re} \left[_2F_1(-\frac{1}{2},\frac{3}{2},3,\frac{2 u_0}{u})\right] \; ,
    \label{eq:full_prod_spec}
\end{align}
where $u_0 = m_ng D/2V$. We can take the simpler expression Eq.\eqref{eq:extractable_prod_spec} for extractable UCN with the understanding that this is to be replaced by the full spectrum Eq.\eqref{eq:full_prod_spec}, which is specific to a particular trap geometry, for \textit{in-situ} calculations.
Allowing for a finite-duration accumulation phase lasting a time $t_a$, followed by a finite-duration storage phase for $t_h$ without further UCN production, the corresponding stored spectra are then
\begin{align}
    \frac{dN(t_a,t_a)}{du} 
    &= \frac{3CV^\frac{3}{2}}{2}\mathcal{V} \cdot \tau_\text{acc}(u) \sqrt{u-u_0} \cdot e^{-\frac{t_h}{\tau_\text{hold}(u)}} \left( 1- e^{-\frac{t_a}{\tau_\text{acc}
    (u)}}\right) \notag \\
    \frac{dN(t_a,t_a)}{du} 
    &= \frac{3CV^\frac{3}{2}}{2}\mathcal{V} \cdot \tau_\text{acc}(u) \sqrt{u} \cdot \text{Re} \left[_2F_1(-\frac{1}{2},\frac{3}{2},3,\frac{2 u_0}{u})\right] \cdot e^{-\frac{t_h}{\tau_\text{hold}(u)}} \left( 1- e^{-\frac{t_a}{\tau_\text{acc}(u)}}\right) \; .
    \label{eq:full_spec}
\end{align}
The number of surviving, extractable UCN is obtained by integrating from $u_\text{min}=E_\text{min}/V$ to $u_\text{max}=E_\text{max}/V$.
The (distinguishable) fundamental parameters of the source then include 
\begin{align}
    \left\{ \; \frac{3CV^\frac{3}{2}}{2}\mathcal{V},u_0,u_\text{min},u_\text{max},\frac{1}{\Lambda}\sqrt{\frac{2V}{m_n}}, f_\text{acc},  \Gamma'_{\text{acc}},f_\text{hold},  \Gamma'_{\text{hold}} \; \right\} \; ,
\label{eq:preparas}
\end{align}
where subscripts distinguish that, e.g., the loss parameters $f_\text{acc}$ and $f_\text{hold}$ may be different during accumulation and storage.

For this proof-of-principle study, we proceed on the assumption that the loss mechanisms during accumulation and holding are identical.
As shown above, values for the first three parameters can be calculated \textit{a priori} on firm physical grounds, while $u_\text{max}=1$ holds for any storage experiment with $t_h$ longer than a few seconds.
We therefore focus on three parameters:
\begin{align}
    \left\{ \; \delta, f, \Gamma' \; \right\} 
    \qquad \text{with} \qquad 
   \delta &= \frac{1}{\Lambda}\sqrt{\frac{2V}{m_n}} \notag \\
   f &= f_\text{acc} \equiv f_\text{hold} \notag \\
   \Gamma^\prime
   &= \Gamma'_{\text{acc}}
   \equiv \Gamma'_{\text{hold}} \; ,
\label{eq:3paras}
\end{align}
where all parameters are assumed to be the same during accumulation and holding. Of greatest interest are the loss parameters $f$ and $\Gamma'$, which determine the maximum achievable UCN density and storage lifetime for real experiments.
Extracting these values also provides diagnosing power for which processes may limit observed performance in practice.
Taking the neutron lifetime as given, the sum of remaining energy-independent losses can be constrained by measurement to establish, e.g., bounds on residual $^3$He contamination of the converter. 
%

\section{Vertical time-of-flight: simulation and measurement}
\label{sec:vTOF}

\begin{figure}[b!]
    \centering
    \includegraphics[width=0.9\textwidth]{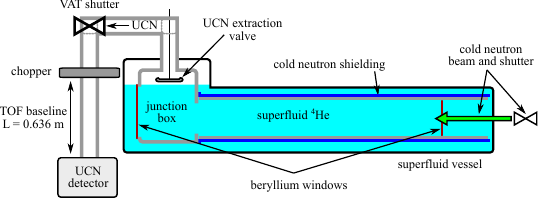}
    \caption{Diagram of SuperSUN~\cite{Degenkolb:2025fbc} for the vertical time-of-flight configuration, with gravity vertical in the plane of the page.}
    \label{fig:supersundiag}
\end{figure}

\begin{figure}[b!]
    \centering
    \includegraphics[width=0.85\textwidth]{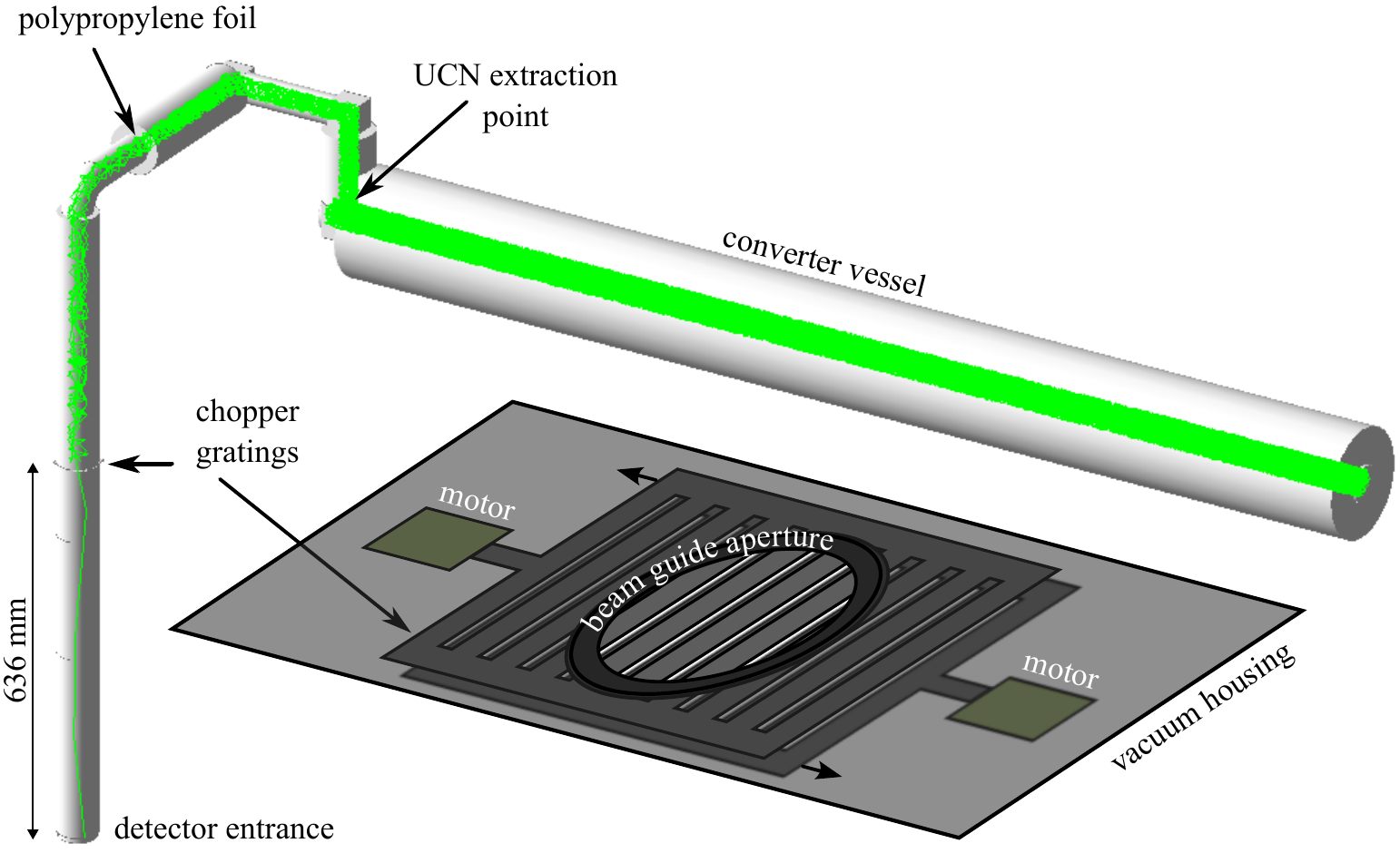}
    \caption{Simulation geometry and representative particle trajectories for the vertical time-of-flight configuration.
    }
    \label{fig:supersundiag2} 
\end{figure}

After accumulation and storage periods, opening the UCN extraction valve releases the stored ensemble for extraction by diffusion as described in Sec.~\ref{sec:ucn}.
UCN leave the helium and exit the converter volume into a vertical guide, passing through two $90^\circ$ bends (see Figs.~\ref{fig:supersundiag} and \ref{fig:supersundiag2}), before falling down to a detector. 
By falling, the UCN gain sufficient kinetic energy to pass the aluminum entrance foil of the detector (optical potential $\sim54\,$neV) with high probability.

\begin{figure}[t] 
    \includegraphics[width=0.395\textwidth]{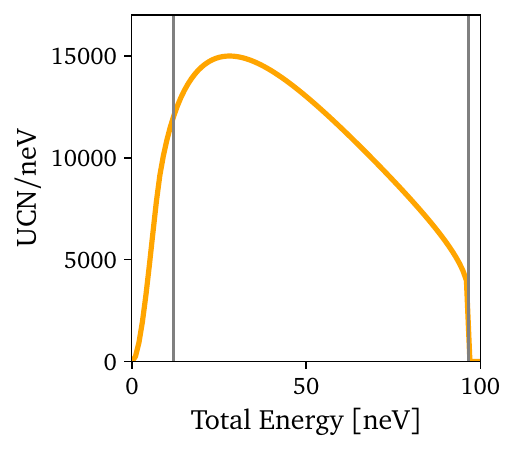}
    \includegraphics[width=0.595\textwidth]{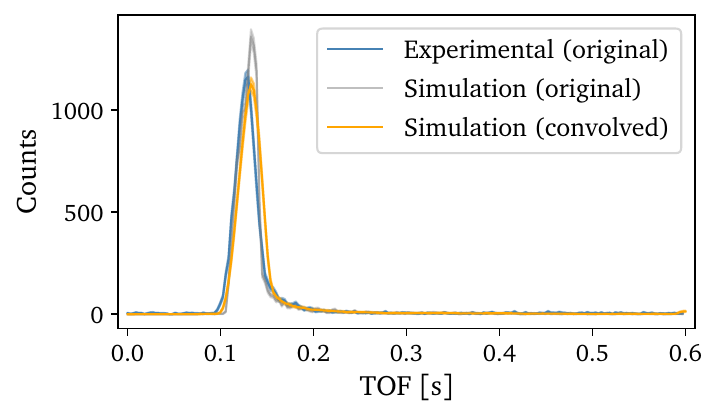}
    \caption{\textbf{Left:} Initial spectrum for $(t_a,t_h) = (1000\,\text{s},100\,\text{s})$. \textbf{Right:} Projected TOF spectrum showing a comparison of experimental data to simulation with and without forward convolution of the chopper transmission function. The pale bands represent $1\,\sigma$ counting statistics errors.}
    \label{fig:deconv_compare}
\end{figure}

The energy spectrum of the extracted population is partially characterised through a vertical time-of-flight (vTOF) measurement, with example data from both measurement and simulation shown in Fig.s~\ref{fig:moneyplot}, \ref{fig:deconvpoor} and \ref{fig:deconv_compare}.
In this work, we use vTOF to refer to the measurement type and TOF to refer to the measured quantity: the observed time-of-flight.

Time-of-flight experiments in general (and vTOF in particular) resolve only one component of the UCN velocity, with uncertainty depending on many factors such as collimation and non-specular reflections.
A chopper cyclically blocks and lets through UCN, see Fig.~\ref{fig:supersundiag2}, with a duty factor around $3\%$ such that each brief open period is followed by a much longer closed period when UCN are blocked from passing.
This defines a series of \textit{frames}, or time windows following each opening, during which any detected UCN are assumed to have passed the chopper only during the most recent opening.
Leakage through the closed chopper appears as a nearly constant background within each frame, proportional to the incident UCN flux. We perform a timing-insensitive correction by subtracting a mean value for each frame, evaluated within a TOF range where no UCN are expected to arrive from the previous chopper pulse.

The moment the chopper opens is recorded electronically using a $5$V transistor-transistor logic (TTL) signal. 
This marks a timestamp that serves as the zero reference for TOF, up to a possible offset that requires calibration.
The open chopper allows UCN to fall through for a short period of time, whereupon they drop for a known length (the TOF baseline), and are then recorded at the detector with a second timestamp.
For a given open-and-close cycle, the TOF for each neutron is given by the detection event's timestamp minus the last TTL timestamp, appropriately corrected for any offset between the TTL signal and the physical opening time.
We add this offset to the TTL timestamp before calculating the TOF.

However, this offset correction is just a first approximation: in practice, the chopper does not open instantaneously but has a time-varying geometrical aperture whose fractional coverage of the entire beam is described by a \textit{chopper transmission function}.
This chopper transmission function was optically measured for the cycle speed used in experimental vTOF measurements, and can be seen in Fig.\ref{fig:deconvpoor} on the left, wherein the time interval between the \emph{left} edge of TTL signal and the midpoint of the symmetrical chopper function defines the chopper offset, measured as here as $21.8\,$ms. 
Fully accounting for the influence of the chopper transmission function on measured spectra is not trivial: in most cases it is simpler to perform a forward convolution on data from a simulation or calculation.
We briefly discuss the associated problem of \textit{deconvolution} at the end of this section.

\subsubsection*{GEANT4 UCN simulations}

Our simulations employ GEANT4~\cite{GEANT4:2002zbu} for particle tracking, adapted  as GEANT4UCN~\cite{ATCHISON2005513} to use ray-tracing for UCN. For this work we have upgraded GEANT4UCN to GEANT4-11-03.2, further extended it, and compared it to other UCN simulation software and analytic models.

A ray-tracing fourth-order Runge-Kutta algorithm (RK4) integrates particle trajectories and provides wall-interaction coordinates, benefiting from all GEANT4 features. This includes, critically for our case, gravity. 
We choose the tracking parameters for integration precision (determining particle position), and the error tolerance for identifying material interfaces (regulating unphysical trajectories that may miss small features) small enough to have no influence on the simulated spectra.
Three foil-type UCN optical components with a physical thickness of $100\,$\textmu m or less are simulated using effective analytical calculations, as described below, rather than adapting the stepping algorithm.
Within these requirements, we allow for maximum step sizes up to $20\,$mm that could be realized, e.g., for straight track segments.

Wall interaction losses follow Eq.\eqref{eq:full_loss}, with parameters encoded as material properties for both the real and imaginary parts of the neutron optical potential, and for the probability of nonspecular reflections.
No approximation or angle-average is performed in the simulation of wall-reflection losses, i.e., the full analytic equation is used at each wall interaction.
Specular reflections are calculated using the surface-normal vector at the interaction point, with transmission and reflection angles following Snell's law.
Non-specular reflections employ a simple $\cos\theta$ distribution around the surface normal, and are implemented with a probability defined independently for each material, e.g., $1$ for very rough surfaces.

Neutron-optical potentials and loss parameters are implemented using either measured or nominal values for all materials.
In ambiguous cases we use the more conservative values, like $f=5\times10^{-5}$ for CYTOP and beryllium at low temperature.
The liquid helium environment of the converter trap is simulated by a shift of the neutron optical potential, Sec.~\ref{sec:UCN_prod}, accounting for boosts and refraction at the interface to other materials.
This is particularly important at the exit of the converter, when UCN leave the helium and enter the first vertically oriented extraction guide.
Importantly, the presence of helium affects the potential step $\Delta U$ that determines reflection and loss probabilities at the converter walls.
We do not consider any losses due to residual gas in the evacuated extraction system; these would present as an additional, energy-dependent, loss for extracted UCN.

Simulations can easily vary the neutron optical potential (including loss), as well as the non-specularity of guide-wall reflections, or the size of mechanical gaps.
Other quantities can be varied (albeit with some effort) both experimentally, and in simulation.
These include the range of travel for opening the extraction valve, the presence or absence of a thin polypropylene foil in the horizontal extraction guide, and the height of drop before and after the chopper along the vertical path to the detector.
In principle, some or all of these can be used for SBI, but we fix them here to nominal values for all simulations and focus on varying $f$ and $\Gamma'$.
In particular the entire system of UCN extraction guides was chosen to have a neutron optical potential of $183\,$neV, a loss factor $f=0.0005$, and a diffuse reflection probability of $0.04$.

We set the neutron $\beta$-decay lifetime to $878\,$s throughout, leading to a constant and universal energy-independent loss rate for all UCN via this channel.
Additional energy-independent losses are implemented only within the converter, by including an additional partial lifetime that can remove particles only within the helium-filled elements.
This is an effective model for neutron capture on $^3$He impurities, allowing us to interpret extracted values of $\Gamma'$ in terms of a limit on $^3$He contamination.

Guide and trap geometries are defined via primitive elements within the simulation, using analytical definitions such that curved surfaces can be simulated without approximation.
We perform the calculations in the spatial domain, with access to time-domain information provided by dedicated code.
For storage experiments the ray-tracing algorithm is adapted to provide pre-defined steps in time, with per-mille absolute precision.
This is necessary to simulate and interpret the time evolution of stored UCN spectra, validating, e.g., that simulations properly capture the time-evolution of ensemble properties.
Particle properties are then recorded at each predefined time step, allowing to track the ensemble evolution through time down to the particle level.
The trap closure at the extraction outlet is approximated, for storage-phase simulations, by a flat reflective surface.

Neutrons are generated homogeneously and with isotropic momenta in the trap volume, using distributions drawn from Eq.\eqref{eq:full_spec} with a particular choice of $t_a$, $t_h$, $f$, and $\Gamma'$.
A few-mm buffer is left at the trap boundary, to prevent initialization errors arising from wall interactions.
The initial distribution rapidly relaxes to mechanical equilibrium, assisted by non-specular reflections with unit probability on the end closures of the converter tube (the cylindrical wall being fully specular).
Fundamental tests were performed, including the relaxation to analytically calculated density distributions in gravity and time-evolution of stored spectra consistent with Eq.\eqref{eq:full_spec}.
These tests also confirmed the precision of the ray-tracing algorithm, and conservation of total energy at the particle level.
At the ensemble level, kinetic energy evolves strongly in time during the approach to mechanical equilibrium, while later both kinetic and total energy can evolve due to energy-dependent loss and extraction.

Since the initial spectra are calculated for total energy, and the total energy distribution at finite $t_a$ and $t_h$ matches the result of first-principles simulations, it is not necessary to simulate UCN storage from the actual moment of production.
Instead, spectra for a given $t_a$ and $t_h$ are calculated using Eq.\eqref{eq:full_spec}, eliminating the need to simulate accumulation and storage sequences that may last thousands of seconds.
Spectrum lists are provided with $1\,$neV resolution and appropriate normalization, and then linearly interpolated within GEANT4 to construct a random sample for simulation.

\subsubsection*{Time-of-flight}

We first establish some notation: we call the chopper period $T$, the open interval during which UCN can pass through it $\Delta T$, the chopper offset $t_{\text{offset}}$, the raw timestamps $t$, the TOF $\tau$, the TOF baseline or distance from chopper to detector $L$, the chopper transmission function $f(t,\tau)$, and the TOF-dependent transmission function through aluminum (here, AlMg3) is $p(\tau)$.

Summarising the approach of Ref.~\cite{NeuVTOF}, the mathematical TOF spectrum $s(\tau, t)$ is defined such that $s(\tau, t)d\tau dt$ is the number of neutrons incident on the chopper between times $t$ and $t+dt$ with times-of-flight $\tau$ and $\tau+d\tau$. The chopper transmission function $0 < f(\tau,t)<1$ is the probability that a neutron with TOF $\tau$, incident on the chopper at time $t$, is transmitted. Calling $b(t)$ the background rate, the counting rate $r(t)$ can be expressed as
\begin{align} 
    r(t) = b(t) + \int{dt' f(t-t', t') p(t-t') s(t-t',t') } \; .
\label{eqn:ratedef1}
\end{align}
We drop the background term, which is accounted for by subtraction of the mean off-signal UCN rate, and neglect $\tau$-dependence of the chopper function. 
Noting that $f(\tau, t)$ for a chopper cycle at time $t=i\cdot T$, ($i=0,1,2\dots$) is only non-zero within $-\Delta T/2 < t < \Delta T/2$, and absorbing the detection efficiency into $s(\tau,t)$, we get
\begin{align}
    r(t) =  \int^{\Delta T/2}_{-\Delta T/2}{dt' f(t') s(t-t',t') } \; .
\end{align}

The time-of-flight apparatus used for experimental measurements consists of a two-grating linear chopper~\cite{Bison:2023aws}, illustrated in Fig.~\ref{fig:supersundiag2} and used in the vTOF configuration~\cite{NeuVTOF}.
For vTOF simulations, the UCN extraction outlet is left open, with a displaced plug mimicking the real extraction valve.
UCN are allowed to enter the guide system, and stochastically explore it up to the detector.
This configuration is also illustrated conceptually in Fig.~\ref{fig:supersundiag}, and is simulated with detailed geometry and materials corresponding to the extraction system used for measurements performed at SuperSUN~\cite{supersundata}.

The titanium grating material leads to small but finite reflection probabilities for low UCN momenta, which is accounted for in simulation by a negative neutron optical potential of $V=-49\,$neV.
Neutrons entering the titanium are lost.
While the neutron guides upstream of the chopper are $50\,$mm inner-diameter stainless steel tubes, the vertical flight tube is an $81\,$mm diameter standard guide section commonly used for this purpose at the ILL's workhorse UCN source PF2.
Due to moving parts near the gratings, an escape channel for UCN to leave the guide system exists in simulation and measurements, with roughly $3\%$ loss in most simulated configurations, with some dependence on the transverse velocity spectrum.

Our simulations consider a static chopper, neglecting deconvolution effects while still capturing some aspects of velocity-dependent UCN transmission. We calculate TOF from the interval between passing the chopper gratings and entering the detector.
At the vertical position of the chopper blades, a timestamp is recorded indicating the start of TOF.
On absorption in the boron carbide detection layer at the bottom of the flight tube, a second timestamp is recorded in addition to other particle parameters.
This is used to compute the raw TOF, relative to the time of passing the chopper.

The detector used for actual measurements was a custom, modified CASCADE-U 100 detector that is approximated in our simulation by a $100\,$\textmu m AlMg3 foil with by a back-side boron carbide coating.
The aluminum foil transmission is also calculated analytically, including non-normal incidence and surface reflections for $V\approx 54\,$neV, and using the measured properties of AlMg3 foils for velocity-dependent neutron absorption~\cite{ATCHISON2009144}.
Our simulations use a boron carbide thickness of $250\,$nm, assuming $^{10}$B$_4$C with $96\%$ enrichment.

Also included in the simulation is a polypropylene vacuum-separation foil upstream of the chopper~\cite{babin2019caracterisation}, whose negative optical potential is used to calculate an energy-dependent effective transmission function.
This transmission function, which also accounts for refraction and angle-of-incidence, is computed analytically at each interaction.
%

\begin{figure}[t]
    \centering
    \includegraphics[width=0.95\textwidth]{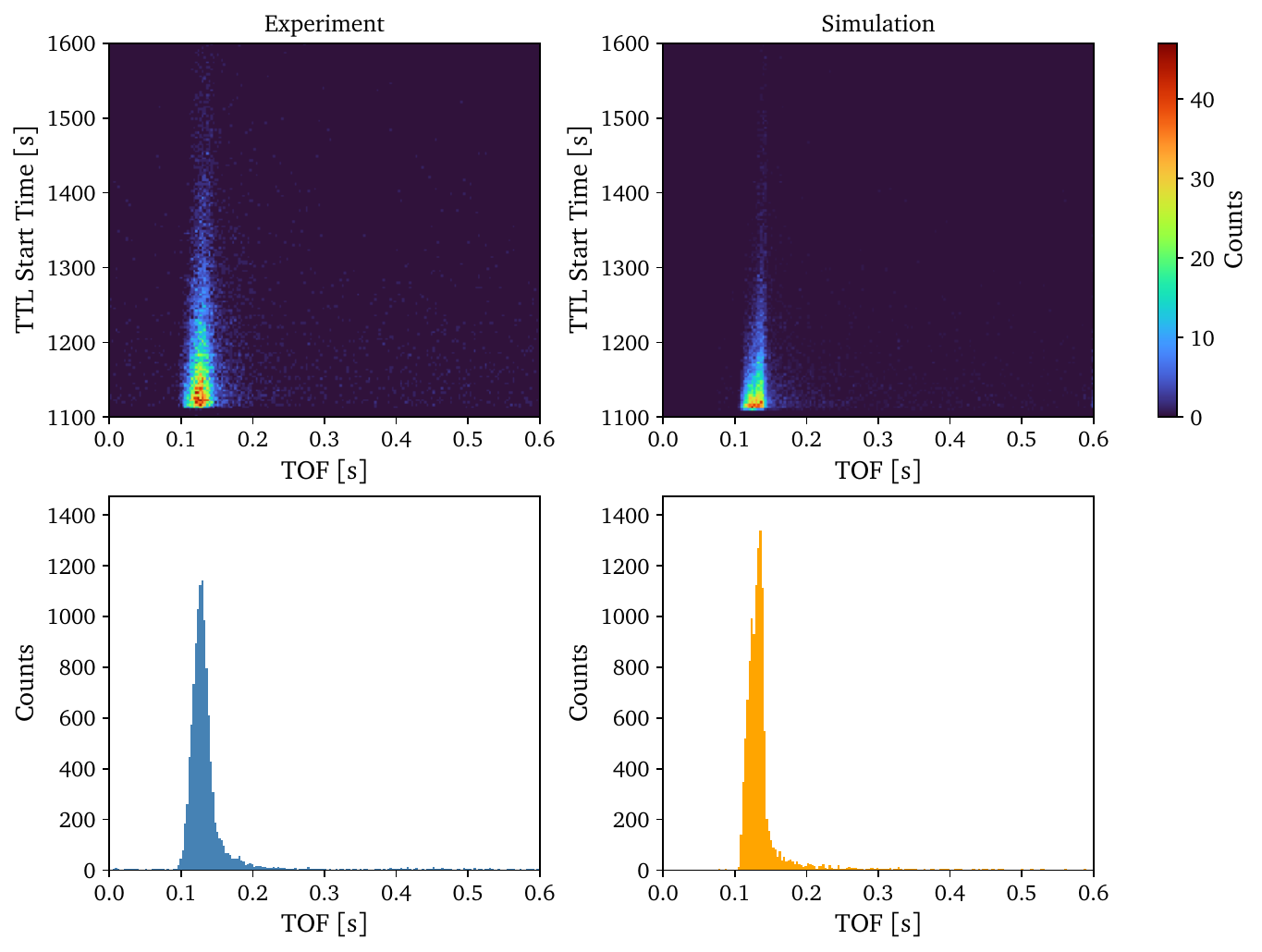}
    \caption{Comparison of measured data (left) and simulation (right) for $(t_a,t_h) = (1000\,\text{s},100\,\text{s})$. Lower plots project the spectrum onto a common TOF ($\tau$) axis. The small discrepancies are partially explained by the static chopper in simulation, i.e., neglecting deconvolution or dynamical effects. For deconvolution see Fig.s~\ref{fig:deconvpoor} and \ref{fig:deconv_compare}, and discussion below. 
}
    \label{fig:moneyplot}
\end{figure}

\subsubsection*{Measurements and experimental vTOF data}

In practice experimental TOF spectra are \emph{estimated} by measuring the number of counts in a time bin, within a chopper frame, rather than the rate.
The observed counts in the $j^{\text{th}}$ bin of width $\Delta \tau$ are:
\begin{align}
    c_j = \int^{t_j+\Delta \tau}_{t_j} r(t) dt \; .
\end{align}
Now considering successive UCN pulses, spaced by the chopper period $T$ such that $T_i = i \cdot T$,
\begin{align} \label{eqn:binsdef}
    c_j = \int^{t_j+\Delta \tau}_{t_j} dt \int dt' \sum_{i=0,1...}^{n} f(t'-i\cdot T) s(t-t',t') \; .
\end{align}
That is, we have an index $i$ for the frames that in real data have length $T=0.6\,$s, and a second index $j$ for each TOF bin within a frame.
Rearranging this sum so that each term is identified by a pair of indices $(i,j)$ defines the matrix $C_{ij}$, which is used for visualization as in Fig.s~\ref{fig:SuperSUN_accumulation} and \ref{fig:moneyplot}.
These pseudo-two-dimensional arrays are also the format for training data in SBI, see Sec.~\ref{sec:inf} below.

Measured UCN statistics are limited by the low duty factor of the chopper, i.e., the short opening interval of $\Delta T\sim 17\,$ms as compared to the chopper period.
To mitigate low statistics, the conventional approach aggregates over chopper frames, rebinning $C_{ij}$ along rows:
\begin{align}
    h_{ij} = \sum_{i'=i}^{i+I-1} c_{i'j}  \; .
\end{align}
The binned (TOF) and aggregated (time) counts matrix is then typically converted to a discrete TOF spectrum as outlined in~\cite{NeuVTOF}: $h_{ij} \rightarrow s_I(\tau_j, t_i)$, where we invert the order of indices as compared to that reference in order to keep TOF on the abscissa (as in most visualisations).
This discrete spectrum is approximately equal to $s(\tau,t)$ for a small $I$.
We emphasize that aggregating over frames projects away information contained in the long-time evolution of the TOF spectrum, which is preserved for our SBI analysis below.

Measurements at SuperSUN have used two choppers, named "Carina" and "Carlos", with chopper offsets $20.2$ms and $21.8$ms respectively.
The configuration we consider here employs Carlos at a $755$mm vertical drop below the center of the horizontal extraction guide, and with a TOF baseline of $L=0.636\,$m.

\subsubsection*{First attempts at deconvolution}

The estimated TOF spectrum outlined above still represents a convolution of the chopper transmission function with the true TOF spectrum, and can be expressed as
\begin{align} \label{eqn:convolution}
s_\text{meas}(t) = (s_\text{true} * f)(t) + \epsilon(t) \; , 
\end{align}
where $s(t)$ could represent either one frame or an aggregation of frames.
Here, $f(t)$ is the chopper transmission function and $\epsilon(t)$ represents noise from backgrounds or Poisson statistics. A deconvolution to the underlying UCN spectrum is challenging; in general, the outcome is non-unique, sensitive to statistical fluctuations, and prone to amplifying noise into large unphysical oscillations (ringing) in the reconstructed signal. 

\begin{figure}[t] 
    \includegraphics[width=0.495\textwidth]{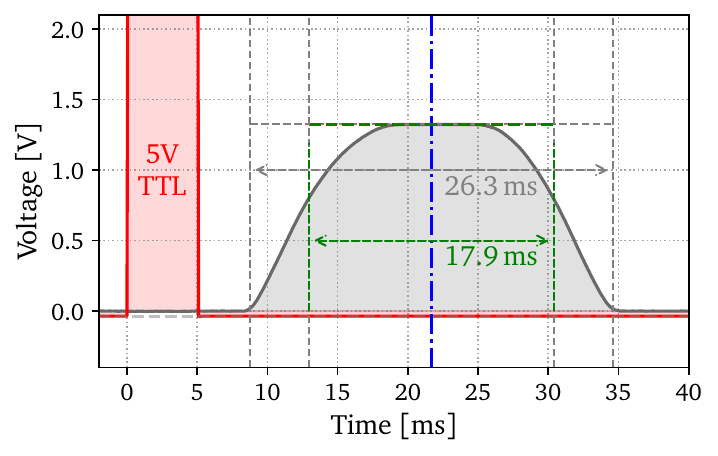}
    \includegraphics[width=0.495\textwidth]{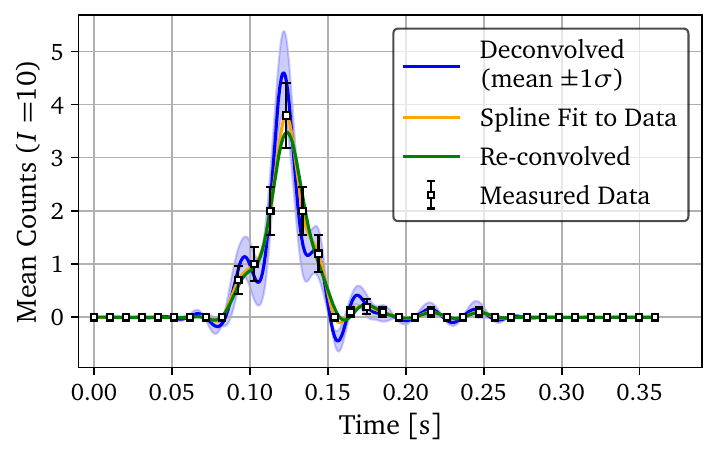}
    \caption{\textbf{Left:} chopper transmission function used for deconvolution, provided by the ILL and based on optical measurements. The grey dashed box defines $\Delta T$. A green dashed box defines the \emph{effective opening time}, by matching a rectangular chopper function to the integral of the true one. \textbf{Right:} Deconvolution using Richardson-Lucy reconstruction, with a cubic spline based seed spectrum, for an accumulation mode TOF slice with low statistics. The spectrum is normalised to standardize fitting parameters.}
    \label{fig:deconvpoor}
\end{figure}

While the noise parameter fundamentally limits deconvolution by fast Fourier transform, smooth functional models provide a more stable basis. Iterative maximum likelihood estimators, such as the Richardson-Lucy~\cite{Richardson:72,1974AJ.....79..745L} algorithm with the update rule
\begin{align} \label{eqn:RL}
s_{k+1}(t) = s_{k} (t) \cdot \left [ \frac{s_\text{meas}(t)}{(s_{k} * f)(t)} * f(-t) \right ] \; ,
\end{align}
which refines the estimate by comparing the forward-convolved trial spectrum with the measured data and applies a multiplicative correction, are a common approach in many fields. In our analysis, the measured spectra are represented by various models fit to binned and unbinned data, such as cubic splines or kernel density estimates, chosen for their smoothness and derivative continuity. This basis reduces susceptibility to noise amplification and provides a flexible yet stable parametrization of the TOF distribution. 

Attempts at RL-based reconstruction of the underlying TOF spectra are shown in Fig.~\ref{fig:deconvpoor}. While applicable to high statistics data with high signal-to-noise, the algorithm remains constrained by the ill-posedness of the problem.

\section{Simulation-based inference}
\label{sec:inf}

The idea of simulation-based or likelihood-free inference is to extract information from a dataset based on a forward simulation. Here the simulations relate our parameters of interest to a data representation and this way replace an explicitly known likelihood.

\subsubsection*{Conditional generative inference}

Simulation-based inference and unfolding are described by a set of four distributions. $p_\text{data}(x)$ describes the reconstructed experimental data. The forward simulation generates $p_\text{sim}(x)$ based on $p_\text{sim}(\theta)$, where $\theta$ can be anything from fundamental model parameters, nuisance parameters, or an intermediate data representation~\cite{Plehn:2022ftl},
\begin{alignat}{9}
  & p_\text{sim}(\theta)
  \quad \xleftrightarrow{\text{\phantom{unfolding inference}}} \quad 
  && p_\text{inf}(\theta)
  \notag \\
  & \hspace*{-9mm} {\scriptstyle p(x|\theta)} \Bigg\downarrow
  && \hspace*{+6mm} \Bigg\uparrow {\scriptstyle p(\theta|x)}
  \notag \\
  & p_\text{sim}(x) 
  \quad \xleftrightarrow{\text{\; forward inference \;}} \quad 
  && p_\text{data}(x)
\label{eq:schematic}
\end{alignat}
The actual forward simulation is then described by the conditional generative probability $p(x|y)$. A standard SBI then compares $p_\text{data}(x)$ with $p_\text{sim}(x)$, either as complete datasets or using summary statistics. For a given inference task the optimal observable or score ensures that the analysis is optimal.

The forward simulation can be inverted using Bayes' theorem to define
\begin{align}
  p(\theta|x) = \frac{p(x|\theta) p(\theta)}{p(x)} \; ,
\end{align}
at the expense of introducing a prior $p(\theta)$. With this second conditional probability, we can define
\begin{align}
  p_\text{inf}(\theta) = \int dx p(\theta|x) p(x) \; .
\end{align}
Depending on the task, we can view $p_\text{inf}(\theta)$ as a simplified data representation that then needs to be compared to $p_\text{sim}(\theta)$. Alternatively, we can view it as a multi-dimensional posterior of the parameters $\theta$.

Technically, this inference relies on our ability to encode the two conditional probabilities. Conditional generative networks are a perfect solution, trained on simulated paired instances $(x,\theta)$ over the joint distribution. We modify the original conditional normalizing flow or cINN encoding $p(\theta|x)$~\cite{Bieringer:2020tnw} by a more modern conditional CFM generator with rational quadratic splines~\cite{Favaro:2025psi}.

We pre-process the input parameters, specifically for $(f, \Gamma')$ we apply a log transformation and then standardize it. For $(t_a, t_h)$, we only standardize the data. The conditional input for the posterior generator is a binned summary statistic of the number of observed neutrons, concatenated with the untransformed parameters. The subnetwork which predicts the parameters of the rational quadratic splines is a fully-connected neural network with ReLU activation. For training we use the AdamW optimizer with default $\beta$ and a cosine-annealing learning rate scheduler. We train the network for 40k iterations using 90\% of the dataset, keeping the remaining 10\% for validation.

\subsubsection*{Benchmark}
\label{sec:inf_toy}

\begin{figure}[t]
    \includegraphics[width=0.495\textwidth, page=1]{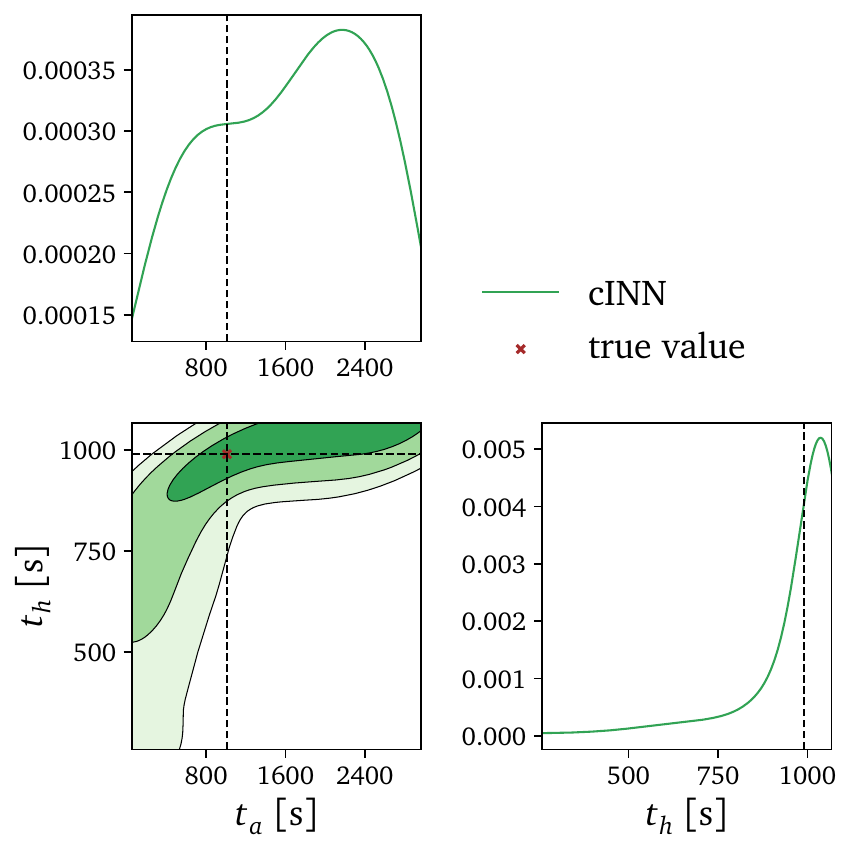} 
    \includegraphics[width=0.495\textwidth]{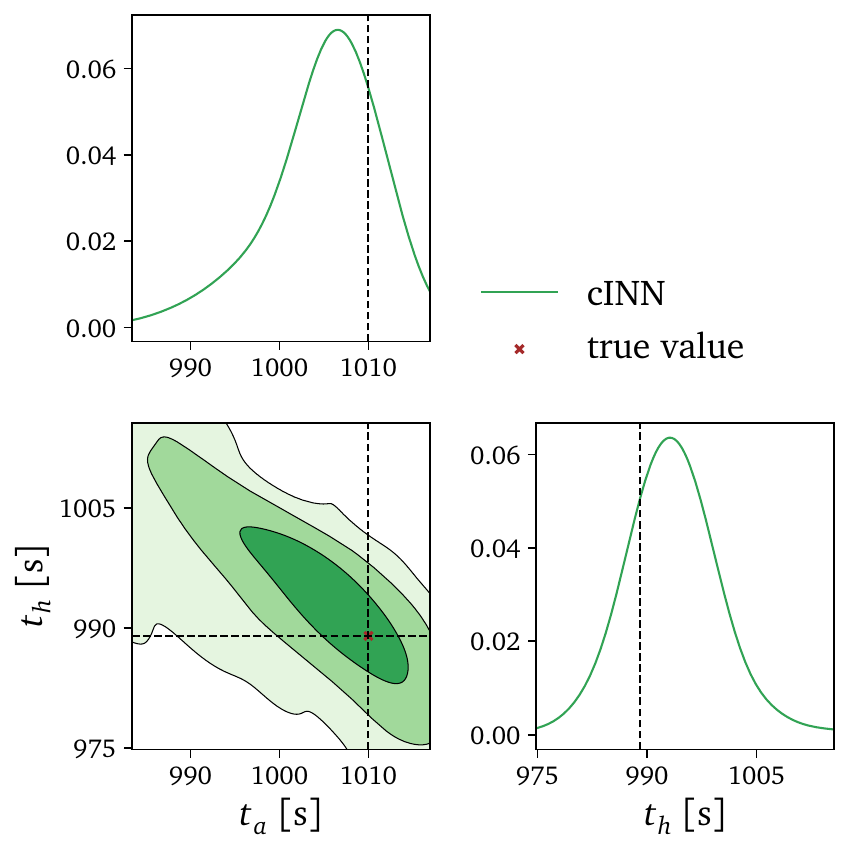} 
    \caption{Neural posteriors for $(t_a, t_h) = (1010\,\text{s}, 989\,\text{s})$. \textbf{Left:} using a different reference time for each simulation. \textbf{Right:} using a global reference time.}
    \label{fig:tA_tH_test}
\end{figure}

To test our method and gain control over our parametrization of physics effects, we consider the toy problem of inferring experimentally defined parameters which have relatively simple implications at the detector level. We select the accumulation and storage times as our toy pair,
\begin{align}
 (t_a, t_h) \; .
\end{align}
Our simulated training dataset contains $\sim$9k simulations with accumulation and storage times uniformly distributed over $[1\,\text{s}, 3000\,\text{s}]$. We start by observing TOF spectra at different times without taking into account a global time delay, i.e., without providing information about the time elapsed before starting to extract UCN.
The conditional information used in the neural network is a summary
statistic of the particles observed at detector level.
We create a 2d-histogram from the observed particles, as in Fig.~\ref{fig:SuperSUN_accumulation}, using Eq.\eqref{eq:full_spec} to fix the relative normalization between simulations.
The histogram contains 20 bins along each dimension, linearly spaced in 
$t\in(0\,\text{s}, 500\,\text{s})$ and $\tau_j \in (0.1\,\text{s}, 0.3\,\text{s})$.

\begin{figure}[t]
    \includegraphics[width=0.495\textwidth]{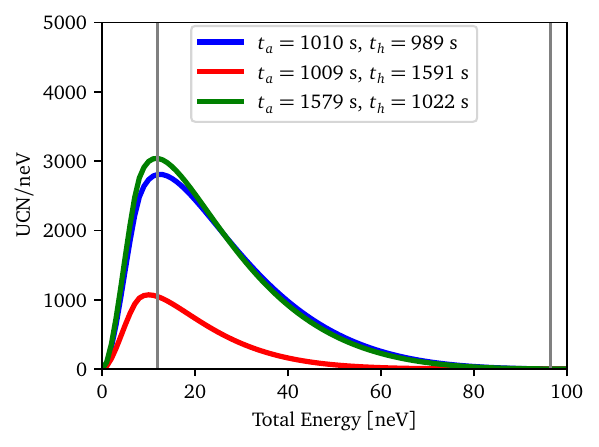}
    \includegraphics[width=0.495\textwidth]{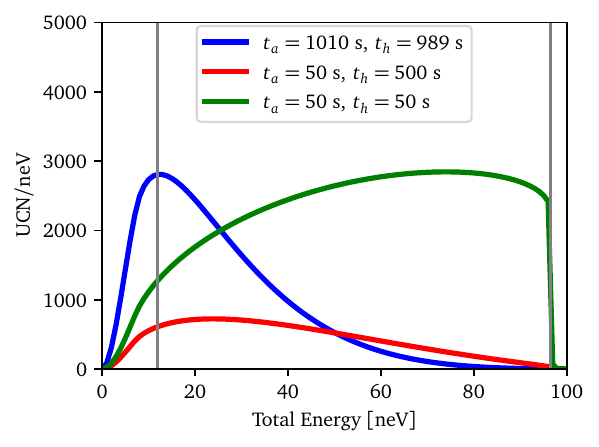}
    \caption{\textbf{Left:} spectra at beginning of UCN extraction. 
    \textbf{Right:} spectra with much shorter $(t_a,t_h)$.
    }
    \label{fig:ta_th_spectra}
\end{figure}

The left panel of Fig.~\ref{fig:tA_tH_test} shows the learned posterior for a test point $(t_a, t_h) = (1010\,\text{s}, 989\,\text{s})$, smoothed with a Gaussian KDE and in terms of 1, 2 and 3~$\sigma$ confidence levels. The probability density is roughly flat in $t_a$ at fixed $t_h$, meaning that for this fixed $t_h$, there is limited constraining power in the accumulation time. 
The physical reason for this is twofold.
On the one hand, Eq.\eqref{eq:full_spec} shows that for sufficiently long $t_a$, UCN produced in each (storable) energy class can reach saturation -- such that further increases of $t_a$ no longer change the spectrum.
On the other hand, for sufficiently long $t_h$, the \textit{shape} of the surviving spectrum becomes relatively insensitive to $t_a$, although changes of $t_a$ do influence its normalization via the total amount of UCN produced.
This effect depends on the relative importance of energy-dependent losses as compared to energy-independent ones.
Because the total loss rate for any given energy is bounded below by $\beta$ decay, the UCN which remain at long $t_h$ are primarily those of such low energy that universal (energy-independent) loss processes dominate.
The corresponding spectra are shown in Fig.~\ref{fig:ta_th_spectra}, and represent the situation at the beginning of UCN extraction for the given $(t_a,t_h)$ pair. They show little qualitative difference, especially above the minimum energy for extraction, and are mainly distinguished by their normalization. Those in the right panel with much shorter $(t_a,t_h)$ exhibit stronger variation, with respect to each other and to the test point. This is what the inference picks up correctly.

For the right panel of Fig.~\ref{fig:tA_tH_test} we simulate a background-free environment where particles arrive at the detector at time $t_a + t_h$.
In this case, the time stamp axis is $(0\,\text{s}, 4500\,\text{s})$.
The learned posterior including the complete information on $(t_a, t_h)$ now constrains both parameters tightly. 

\subsubsection*{Inference}

Joint inference of the two loss parameters $(f, \Gamma')$ follows the same procedure. We generate 9k simulations with parameters sampled according to uniform distributions,
\begin{align}
    f\sim \mathcal{U}\left(\frac{10^{-5}}{\pi}, 2\pi\cdot10^{-4}\right) \qquad \text{and} \qquad 
    \Gamma' \sim \mathcal{U}\left({\Gamma'}_\text{min}, {\Gamma'}_\text{max}\right) \;.
\end{align}
We perform the training and the inference in log-space, as we observe a numerically more stable convergence. 
The range for $f$ is motivated by measurements of UCN storage with the $115\,$neV material CYTOP~\cite{Neulinger:2022oue} which forms the limiting wall potential in SuperSUN's converter, and possible additional mechanical gaps or small absorbing areas on the walls.
The loss factor for CYTOP is in the range of a few $\times 10^{-5}$ at low temperatures, with the lowest measurement to date having been performed at just above $10\,$K.
We conservatively estimate a minimum value for $0.6\,$K as $f=1.0\times 10^{-5}$, and include a safety factor of $\pi$ when defining the limits of the parameter range for training.
More likely, other materials or gaps in the converter walls may increase $f$ above the bare value for CYTOP alone.
We assume a maximum $f=2.0\times 10^{-4}$, again extending the range by $\pi$ for training data.
This admits equal, and overly-pessimistic, contributions on the level of $10^{-4}$ from \textit{both} gaps and neutron capture.
A $10^{-4}$ contribution to $f$ from mechanical gaps would be consistent with physical gaps of $0.1\,$mm at the converter's mechanical interfaces, i.e., at the maximum of the intended design tolerance.

For $\Gamma'$ the range is bounded below by neutron decay: we account for the present experimental uncertainty in the neutron lifetime by taking a minimal value $\Gamma'_\text{min} = 1.12\times 10^{-3} \,\text{s}^{-1}$ that corresponds to $\tau_n= 895\,\text{s} $, with no further energy-independent losses.
We do not extend the training range into the unphysical parameter space of longer neutron lifetimes.
The upper limit for $\Gamma'$ is motivated by practical limits on $^3$He contamination in the converter.
A simple relation is employed to determine energy-independent losses from storage in a bulk medium contaminated with $^3$He,
\begin{align}
    \Gamma_{^3\text{He}} =\left|\Gamma' - \left( 878\,\text{s} \right)^{-1}\right| \; , 
\end{align}
where the absolute value provides robustness against floating-point numerical errors.
These are exacerbated by simulations calculating neutron loss in the time domain.
A relative $^3$He fraction above $8\times10^{-11}$, corresponding to $\Gamma' > 0.0031\,\text{s}^{-1}$, is excluded by integral storage measurements already performed at SuperSUN~\cite{Degenkolb:2025fbc}.
However, we again include a safety factor of $\pi$, such that $\Gamma'_\text{max}\approx 0.0098\,\text{s}^{-1}$.

\begin{figure}[t]
    \centering
    \includegraphics[width=0.6\textwidth]{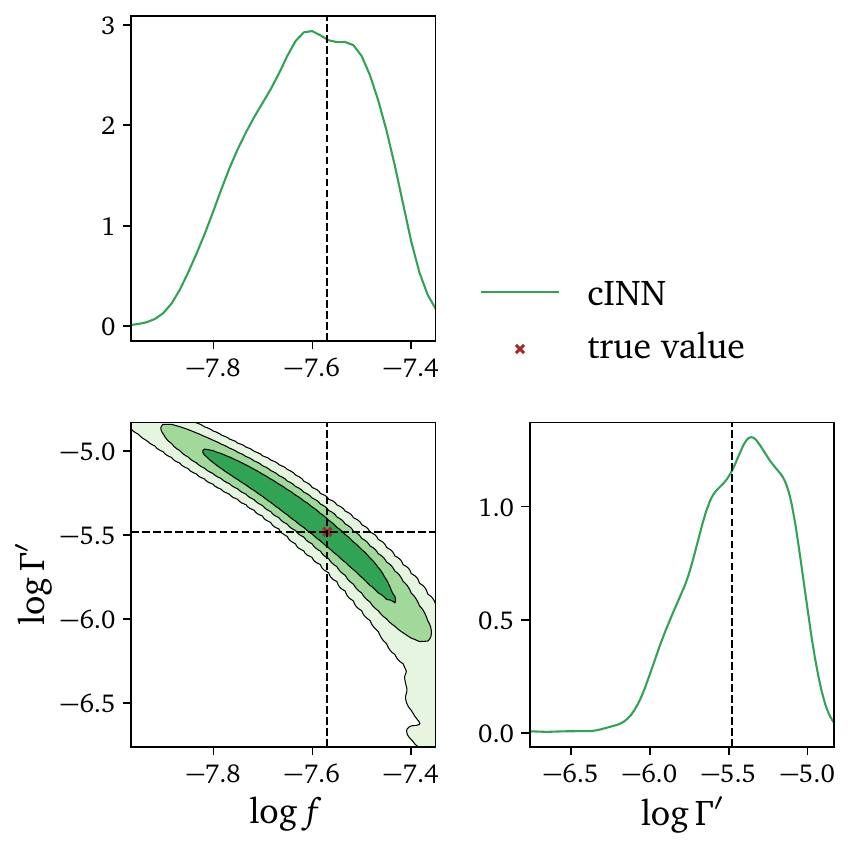} 
    \caption{Neural posteriors for the inference of $(\log f, \log \Gamma')$. The posterior is extracted from a test simulation with parameters $(-7.57,-5.48)$. Contours show 1/2/3$\sigma$ surfaces.}
    \label{fig:f_Gamma}
\end{figure}

In Fig.~\ref{fig:f_Gamma} we show the correlated posterior for $f$ and $\Gamma'$. The main feature, a strong anti-correlation of $f$ and $\Gamma'$ is expected: the presence of a certain amount of surviving UCN, for a given $t_a$ and $t_h$, sets a bound on the time constants for decay which include contributions from both $f$ and $\Gamma'$.
To maintain consistency with observation, increasing $f$ therefore requires decreasing $\Gamma'$, and vice versa.
This condition applies separately for each UCN energy in the surviving population, and for long $t_a$ or $t_h$ the effects discussed above in the toy model tend to relax the constraining power of simulations or measurements.
We have chosen $(t_a,t_h)=(500\,\text{s},100\,\text{s})$ to preserve a mixture of energy-dependent and energy-independent effects while also keeping an experimentally relevant range of parameters.
The posterior is broader in $\Gamma'$, while it constrains a large parameter region in $f$. In particular, we observe a unimodal shape with small values of $f$ largely excluded.
From the marginal distribution and the contour levels, we also observe that the ground-truth value lies in the high-density region of the posterior. While a quantitative coverage study requires further investigation, the shape and the coverage of these posteriors is of great experimental interest.

\section{Outlook}
\label{sec:outlook}

Simulation-based inference is becoming the standard methodology across many physics directions, especially when we want to extract low-dimensional fundamental physics information from complex experimental setups or complex datasets. This development is driven by modern machine learning. The main problem is that SBI is conditional on precise, high-quality simulations.
Specifically, extraction-based UCN measurements suffer from energy-dependent efficiency, which modifies the stored UCN spectrum before detection in ways that are difficult to assess from measured data, and from low statistics which can obscure deviations of simulations from measurement.

The attendant difficulty in interpreting measurements or simulation results arises partly from severe ambiguity in which of several correlated effects may actually dominate observed data trends.
The spectral softening at long counting times can arise from many underlying processes, including effects as simple as the fact that low-energy UCN propagate more slowly through a guide system.
As we have demonstrated, the constraining power of measured or simulated data is low in large parts of the relevant parameter space, but other points do provide constraining power and precise simulations can be exploited to recover the underlying parameter values via SBI.

In integral measurements the arrival time is also related to UCN velocity, but due to correlations and remixing of velocity components in the extraction guides, this is not straightforward to interpret.
This has long presented difficulties in the interpretation of experimental data, since different quantities can be altered in experiment as compared to models or simulations.
Our analysis shows that vTOF together with SBI provides a viable alternative to begin better understanding the underlying physics of UCN storage, in sources or storage volumes that couple to detectors only via complex apparatus.

Since UCN are in general detected only after complex multi-step experimental operations, the field is badly in need of an approach to disentangle correlations and intermediate physics effects between production and detection.
In that context, we have presented a first demonstration of SBI based on precise and reliable vTOF simulations for the UCN source SuperSUN.
We have shown how UCN production and the interactions within the converter volume can be simulated in GEANT4, focusing on a simple model highlighting our inference parameters $(f,\Gamma')$ as being of particular experimental interest -- both for understanding SuperSUN itself, and for building towards more complex experimental setups in which the same loss mechanisms also operate.
Finally, we demonstrated that neural simulation-based inference enables the extraction of full multidimensional posteriors, for parameters which allow to to characterize UCN behavior in terms of underlying physics processes and therefore generalize that understanding to broader situations.
\section*{Acknowledgements}

We are indebted to the SuperSUN-PanEDM collaboration and to the ILL's Nuclear and Particle Physics group for extensive stimulating discussions and support, and are very grateful for a close working relationship with the ILL's SANE group at SuperSUN. We also thank Manuel Schiller for a crucial series of technical recommendations and personalized computing support. This research is supported by the Deutsche Forschungsgemeinschaft (DFG, German Research Foundation) under grant 495942900 \textsl{Search for the Electric Dipole Moment of the Neutron with the PanEDM Experiment}, grant 396021762 --  TRR~257: \textsl{Particle Physics Phenomenology after the  Higgs Discovery}, and through Germany's Excellence Strategy EXC~2181/1 -- 390900948 (the \textsl{Heidelberg STRUCTURES Excellence Cluster}). 

\appendix

\section{Training hyperparameters}
\label{app:hyperparam}
We include in Table~\ref{tab:hyperparams} the details of the neural network training and the hyperparameters of the architecture.
\begin{table}[]
    \centering
    \begin{tabular}{lc}
        \toprule
        Parameter & Value \\ \midrule
        Iterations & 40000 \\
        LR sched. & cosine\\
        Max LR    & $10^{-4}$   \\
        Optimizer & AdamW \\
        $[\beta_1, \beta_2]$ & $[0.9, 0.999]$ \\
        Batch size & 128 \\
        \midrule
        Transformation & RQS \\
        N. of bins & 8 \\
        N. of blocks & 3 \\
        Boundaries & $[-3, 3]$ \\
        Neural network  &  MLP  \\
        N. of layers   & 3  \\
        Hidden channels & 128 \\
        \bottomrule
    \end{tabular}
    \caption{Hyperparameters used to train the cINN, followed by the parametrization of the rational quadratic spline and the parameters of the neural network.}
    \label{tab:hyperparams}
\end{table}
%

\bibliography{tilman,references}
\end{document}